\documentclass[twocolumn,showpacs,superscriptaddress]{revtex4}

\usepackage{graphicx}

\newcommand{\avg}[1]{\langle{#1}\rangle}

\begin{document}

\title[Electrostatics in PBC and Real-space Corrections]
{Electrostatics in Periodic Boundary Conditions and Real-space Corrections}

\author{Ismaila Dabo}
\email{dabo_is@mit.edu}
\affiliation{Department of Materials Science and Engineering, 
Massachusetts Institute of Technology, Cambridge, MA, USA}
\author{Boris Kozinsky}
\affiliation{Department of Physics,
Massachusetts Institute of Technology, Cambridge, MA, USA}
\author{Nicholas E. Singh-Miller}
\affiliation{Department of Materials Science and Engineering,
Massachusetts Institute of Technology, Cambridge, MA, USA}
\author{Nicola Marzari}
\affiliation{Department of Materials Science and Engineering,
Massachusetts Institute of Technology, Cambridge, MA, USA}

\begin{abstract}
We address periodic-image
errors arising from the use of periodic boundary conditions to describe
systems that do not exhibit full three-dimensional periodicity.
The difference between the periodic potential, 
as straightforwardly obtained from a Fourier transform, and the potential 
satisfying any other boundary conditions
can be characterized analytically.
In light of this observation, we present an
efficient real-space method to correct periodic-image errors,
based on a multigrid solver for the potential difference, and
demonstrate that exponential convergence
of the energy with respect to cell size can be achieved
in practical calculations. 
Additionally, we derive rapidly convergent expansions for determining
the Madelung constants of point-charge assemblies in one, two, and
three dimensions.
\end{abstract}


\maketitle

\section{Introduction}


First-principles calculations frequently employ periodic boundary conditions
to predict materials properties. Besides
constituting a natural choice when studying crystalline systems,
periodic boundary conditions allow the use of highly optimized fast
Fourier transform (FFT) algorithms \cite{FrigoJohnson2005,CooleyTukey1965,HeidemanJohnson1985},
which considerably reduce the computational cost
associated with the resolution of electrostatic equations, and allow
an efficient evaluation of electronic kinetic energies and interatomic forces
when used in conjunction with a plane-wave basis set. 
Despite these algorithmic advantages, periodic boundary conditions
require large supercells when studying
aperiodic or partially periodic systems ({\it e.g.}, isolated molecules, polymer chains, and 
slabs) in an effort to minimize spurious electrostatic interactions 
between periodic images \cite{LeslieGillian1985}. Charged systems are
particularly problematic, since conventional 
algorithms automatically enforce charge neutrality by introducing an artificial 
jellium background \cite{LeslieGillian1985}. (Note that the electrostatic energy 
of a charged system exhibiting three-dimensional periodicity is infinite.)
As shown by Makov and Payne, these artifacts induce 
significant errors scaling as $1/L^3$ for the energy of neutral polarized
systems and $1/L$ for that of charged systems,
where $L$ denotes the size of the unit cell \cite{MakovPayne1994}.


In addition to the Makov-Payne asymptotic correction \cite{MakovPayne1994}, 
several schemes have been devised to reduce periodic-image errors.
Barnett and Landman proposed to eliminate periodic-image interactions for
cluster systems by restricting the plane-wave expansions of the wavefunctions 
and of the charge density
to a spherical domain in reciprocal
space \cite{BarnettLandman1993,MarxHutter1995,MarxHutter2000}. A generalization of
this reciprocal-space approach was introduced by Martyna and Tuckerman
\cite{MartynaTuckerman1999}.
The electrostatic-cutoff approach proposed by Jarvis, White, Godby, and Payne
suppresses periodic-image effects by damping the electrostatic 
potential beyond a certain interaction range \cite{JarvisWhite1997}.
The corrective method introduced by Bl{\"o}chl
consists of using atom-centered Gaussian charges 
and  Ewald summation techniques  to
cancel periodic-image interactions \cite{Blochl1995}.
In the local-moment-countercharge (LMCC) method developed by Schultz,
a superposition of Gaussians is employed as a local-moment 
model for calculating the Coulomb potential analytically up to a
certain multipole order, the remaining electrostatic contribution 
being computed using conventional plane-wave techniques \cite{Schultz1999}.
Considering atomic adsorption on neutral slabs, Neugebauer and
Scheffler proposed eliminating the adsorbate-induced
polarization through the introduction of a counteracting planar dipole between slab 
images \cite{NeugebauerScheffler1992}. Refinements of this
method, based on the linear- and planar-average 
approximations proposed by Baldereschi,
Baroni, and Resta \cite{BaldereschiBaroni1988}, 
were subsequently developed \cite{Bengtsson1999,NatanKronik2000,MeyerVanderbilt2001}. 
Extending this approach to
charged surfaces, the prescription of Lozovoi and Alavi relies 
on inserting a Gaussian layer in vacuum to compensate for the excess 
charge and to allow electric-field discontinuities across the layer 
\cite{LozovoiAlavi2003}. 


In this work, we propose an alternative approach for correcting periodic-image 
errors and show that exponential energy convergence with respect to
cell size can be obtained at tractable computational cost. 
The approach proceeds by calculating the electrostatic potential
in real space, exploiting the periodic solution of the Poisson equation
computed using inexpensive FFT techniques. 
In the following sections, we first 
discuss and characterize the difference between the open-boundary
electrostatic potential 
and its periodic counterpart, providing a comparative basis for analyzing
the relative accuracy of various corrective schemes. 
Second, we present our correction method 
and assess its performance. Last, we extend the method to the study 
of systems exhibiting one- or two-dimensional periodicity, beyond the conventional
linear- and planar-average approximations.

\section{Comparison of the Open-boundary and Periodic Potentials}

\label{CorrectivePotentialSection}

\subsection{Definition of the Corrective Potential}

\begin{figure*}
\includegraphics[height=10cm]{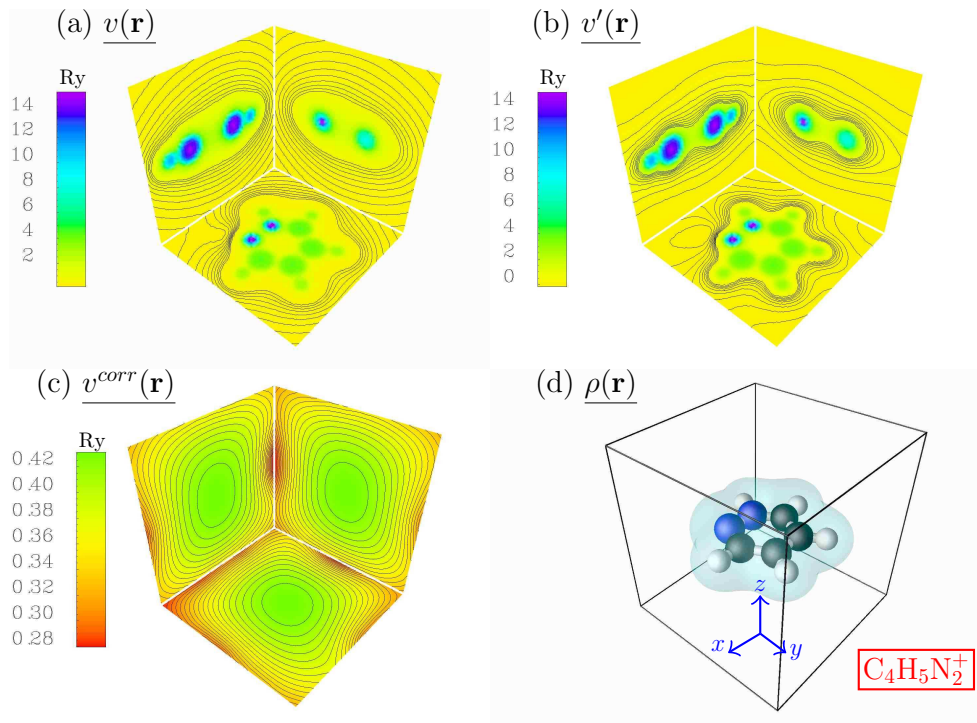}
\caption{(a) Open-boundary electrostatic potential $v$, (b) periodic
electrostatic potential $v'$, and (c) electrostatic-potential correction
$v^{corr}=v-v'$ for a pyridazine cation in a cubic cell of length $L=15$ bohr. The potentials
are plotted in three orthogonal planes $(Oxy)$, $(Oxz)$, and $(Oyz)$ passing
through the center of the cell.
\label{ParabolicCorrection}}
\end{figure*}


The electrostatic potential $v$ generated by a charge distribution $\rho$
satisfies the Poisson equation:
\begin{equation}
\nabla^2 v({\bf r}) = - 4 \pi \rho({\bf r})
\label{CoulDifferential}
\end{equation}
(atomic units are used throughout).
In the absence of an  external electric field, we can
solve Eq. \ref{CoulDifferential} subject to open-boundary conditions
($v({\bf r}) \to 0$ as $|{\bf r}| \to +\infty$). 
As a result,
the electrostatic potential $v$ can be computed
via Coulomb integration:
\begin{equation}
v = \int \frac{\rho({\bf r}')}{| {\bf r} - {\bf r}' |}d{\bf
r}'.
\label{CoulEquation}
\end{equation}
(Although this study focuses on open boundary conditions,
it should be noted that the contribution from an external field ${\bf E}$ can be 
incorporated by adopting the 
asymptotic boundary conditions $v({\bf r}) \to - {\bf E}\cdot{\bf r}$,
which simply adds a term $- {\bf E}\cdot{\bf r}$ to the solution
of the Poisson equation.)
A differential equation similar to Eq. \ref{CoulDifferential}
can be written for the periodic potential $v'$,
keeping in mind that periodic boundary conditions can 
only accommodate a net zero charge (as seen from Gauss' law):
\begin{equation}
\nabla^2 v'({\bf r}) = - 4 \pi (\rho({\bf r}) - \avg{\rho} ).
\label{FFT_Differential}
\end{equation}
As a consequence, the periodic potential can be evaluated
in the reciprocal-space representation as:
\begin{equation}
v'({\bf r}) = \sum_{{\bf g} \neq {\bf 0}} \frac{4\pi}{{\bf g}^2}
\rho({\bf g})e^{i{\bf g \cdot r}},
\label{FFTEquation}
\end{equation}
where we set the arbitrary component $v'({\bf g}={\bf 0})=\avg{v'}$ to zero.


\begin{figure}
\includegraphics[width=8.5cm]{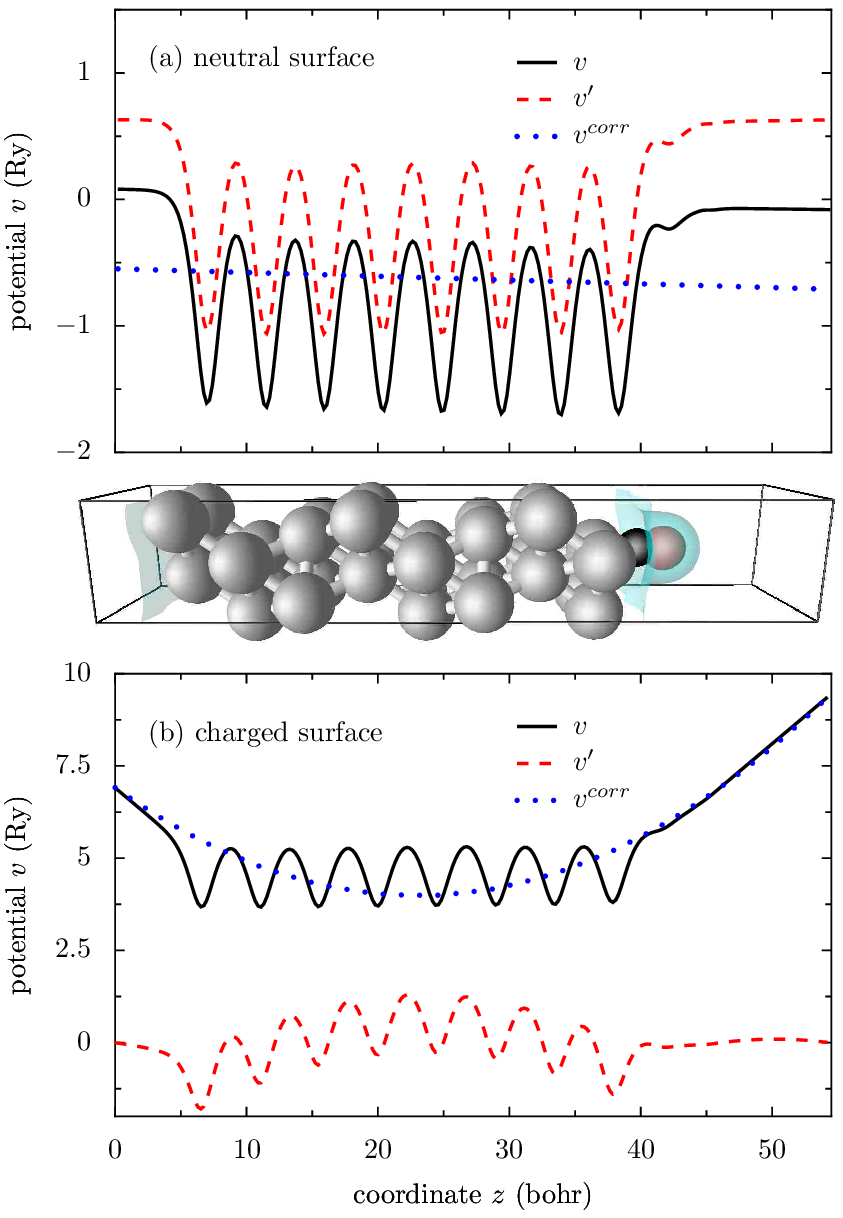}
\caption{Open-boundary electrostatic potential $v$, periodic
potential $v'$, and electrostatic-potential correction
$v^{corr}$
averaged in the $xy$-plane parallel to the surface
for (a) carbon monoxide adsorbed on a neutral platinum slab,
 and (b) carbon monoxide adsorbed on a charged platinum slab.
\label{8ML_Linear}}
\end{figure}

It should be noted that the open-boundary potential $v$ and 
its periodic counterpart $v'$ are distinct. We define
the corrective potential $v^{corr}$ as the difference
$v-v'$. The potential $v^{corr}$ must satisfy:
\begin{equation}
\nabla^2 v^{corr}({\bf r}) = -4 \pi \avg{\rho},
\label{corrDifferential}
\end{equation}
for which we specify Dirichlet boundary conditions: $v^{corr} = v - v'$
at the cell boundaries. (Note that the solution of this
elliptic boundary value problem \cite{Garabedian1964,Ames1992} is uniquely
defined.) 
Eq. \ref{corrDifferential} indicates that the curvature
of the corrective potential is a constant.
It should also be noted that, apart from
the value of the average $\avg{\rho}$, Eq. \ref{corrDifferential} is
independent of the structural details of the charge density $\rho$.
Instead, these details are entirely embedded in the Dirichlet boundary conditions,
which reflect the electrostatic contributions from 
compensating jellium and from the surrounding images.

In order to illustrate the implications of Eq. \ref{corrDifferential},
we consider a pyridazine cation in a periodically repeated cubic cell. 
The open-boundary potential $v$, the
periodic potential $v'$, and the corrective potential
$v^{corr}$ are 
shown in Figure \ref{ParabolicCorrection}.
First,
we observe that the potential $v'$ is shifted down in energy with respect to $v$,
due to the fact that the average $\avg{v'}$ is null by construction.
In addition to this energy shift,
the potential $v'$ is significantly distorted. 
This distortion results from satisfying the
periodicity conditions.
Most importantly, we observe that the corrective potential
$v^{corr}$ varies smoothly over space.
The smooth spatial dependence of $v^{corr}$ contrasts markedly with the strong
variations in $v$ and in $v'$. Performing a polynomial regression, 
we can verify that the potential $v^{corr}$ is quadratic to good
approximation in the proximity of the cell center with
departures from parabolicity restricted to the vicinity
of the periodic boundaries.

To further examine the characteristics of $v^{corr}$,
we consider the adsorption of carbon monoxide 
molecules on neutral and charged platinum slabs.
Following Neugebauer and Scheffler, the electrostatic correction is calculated 
along the $z$-direction within the planar-average approximation (that is,
from the $xy$-average of the charge distribution) 
\cite{BaldereschiBaroni1988}. The 
validity of this approximation is discussed in the last section.
For CO molecules adsorbed on a neutral slab (Figure \ref{8ML_Linear}a),
the periodic potential is shifted up in energy and tilted 
with respect to the open-boundary potential. 
The potential correction is seen to be linear,
in agreement with the analysis of Neugebauer and Scheffler \cite{NeugebauerScheffler1992}.
For CO molecules adsorbed on a slab of surface charge 
$\tilde \sigma$ (Figure \ref{8ML_Linear}b), the real-space 
potential diverges as $4\pi\tilde \sigma|z|$. In this case,
the periodic potential $v'$ undergoes a significant energy downshift, which
decreases the energy of the positively
charged slab. Moreover, we observe that 
$v'$ is significantly curved in the slab region. 
Consistent with these observations and with Eq. \ref{corrDifferential}, 
the corrective potential $v^{corr}$ 
is found to be parabolic everywhere in the unit cell.

\subsection{Quasiparabolic Behavior of the Corrective Potential}

\begin{figure}
\includegraphics{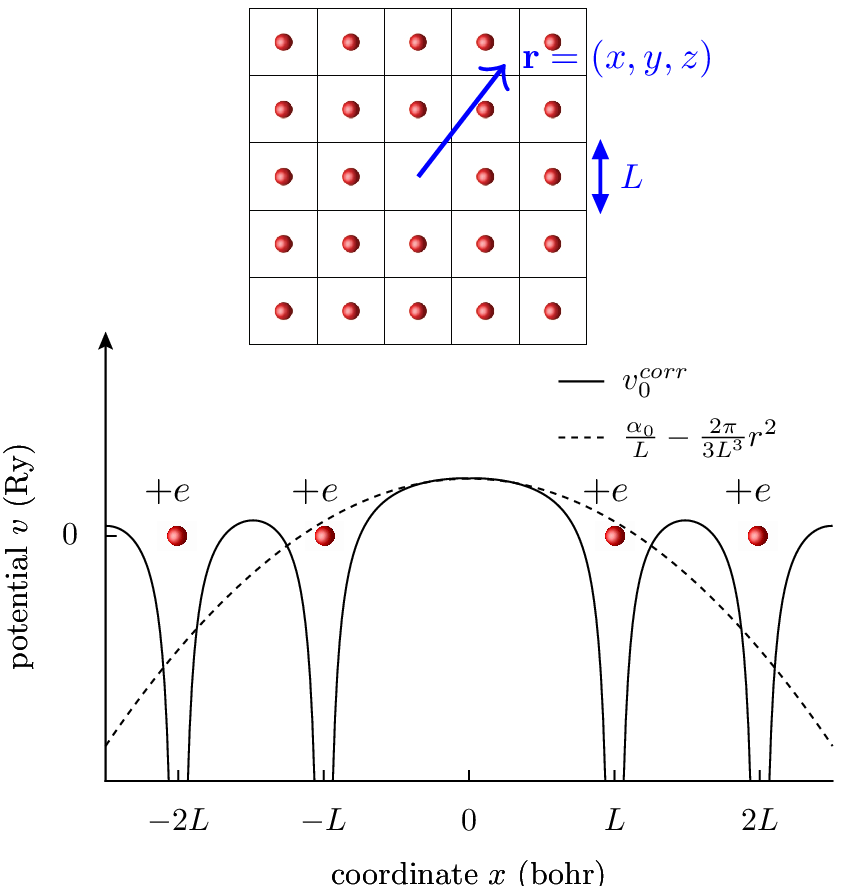}
\caption{Corrective potential $v_0^{corr}$ for a cubic lattice of point charges and
its parabolic approximation in the vicinity of the origin.}
\label{PointChargeCorrection}
\end{figure}


In order to complete the analysis of the corrective potential,
we consider a point charge $q=+e$
in a periodically repeated cubic cell of length $L$, as illustrated in Figure 
\ref{PointChargeCorrection}.
The corrective potential generated by the uniform jellium
and the surrounding point charges is denoted $v^{corr}_0$.
Note that $v^{corr}_0$ cannot be calculated directly as the difference
between the potential of a lattice of point charges $v'_0$ and
the point-charge potential $1/r$ since the representation of a point charge
in reciprocal space requires an infinite number of plane-wave components.
Instead, to obtain $v^{corr}_0$, we can exploit the cubic symmetry 
of the system, writing the corrective potential as:
\begin{eqnarray}
v^{corr}_0({\bf r})&=&v^{corr}_0(r=0) \nonumber \\
&&+ \nabla^2 v^{corr}_0(r=0) \frac{r^2}{6} + O(|{\bf r}|^4).
\label{CorrectionExpansion}
\end{eqnarray}
This parabolic expansion, valid up to third order, confirms that the point-charge
correction $v^{corr}_0$ is almost quadratic in the vicinity of $r=0$. For
noncubic lattices, due to inversion symmetry,
the point-charge corrective potential takes a more general form:
\begin{eqnarray}
v^{corr}_0({\bf r})&=&v^{corr}_0(r=0) \nonumber \\
&&+ \frac{1}{2} \sum_\alpha \frac{\partial^2 v^{corr}_0}{\partial r_\alpha^2}(r=0) r_\alpha^2 + O(|{\bf r}|^4),
\end{eqnarray}
where $(r_\alpha)$ are the coordinates of ${\bf r}$ along the principal axes.
Thus, the corrective potential in a noncubic lattice is also quasiparabolic. 


Turning now to an arbitrary distribution $\rho$,
we can express the electrostatic correction $v^{corr}$ 
by superposition:
\begin{equation}
v^{corr}({\bf r})=\int v^{corr}_0({\bf r}-{\bf r}')\rho({\bf r}')d{\bf r}'.
\end{equation}
As a consequence, defining $r_{max}$ as the distance
beyond which the parabolic expansion (Eq. \ref{CorrectionExpansion})
ceases to be valid, 
the corrective potential $v^{corr}$ can be
considered as nearly parabolic, provided
that the spread of the distribution is tolerably lower
than $r_{max}$.

\subsection{Connection with Existing Schemes}

\begin{figure*}
\includegraphics[width=16cm]{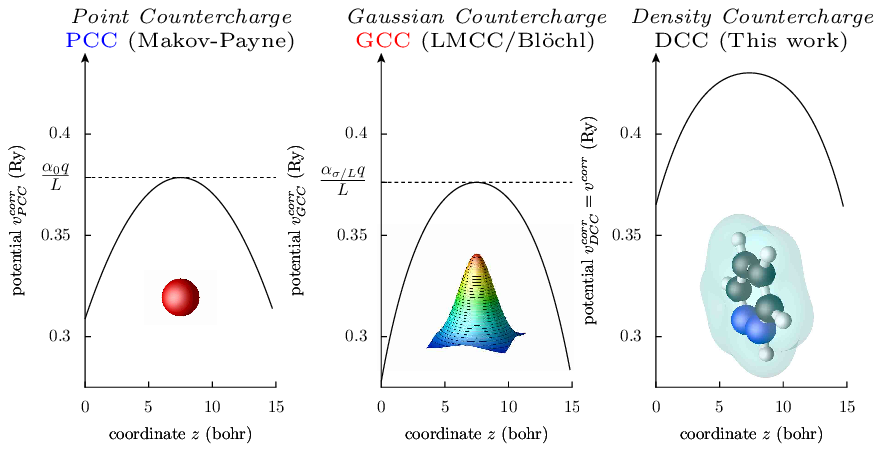}
\caption{Point-countercharge (PCC), Gaussian-countercharge (GCC),
and density-countercharge (DCC) corrective potentials for a pyridazine cation
C$_4$H$_5$N$_2^+$ in a cubic cell of length $L$ = 15 bohr. The corrective potentials
are plotted along the $z$-axis perpendicular to 
the plane of the molecule, as defined in Figure \ref{ParabolicCorrection}.
The PCC and GCC corrections
are calculated up to dipole order. The spread of the Gaussian countercharges is
$\sigma=0.5$ bohr.}
\label{PyridazineCationComparison}
\end{figure*}


Having justified the general characteristics 
of the electrostatic-potential correction, 
we now determine the terms in the expansion of $v^{corr}_0$ 
(Eq. \ref{CorrectionExpansion}). 
The potential at the origin $v^{corr}_0(r=0)$ can be 
written in terms of the Madelung constant $\alpha_0$ \cite{Ziman1972} of a cubic lattice 
of point charges in a compensating jellium background:
\begin{equation}
v^{corr}_0(r=0)=\frac{\alpha_0}{L}.
\end{equation}
(The calculation of the Madelung constant of a jellium-neutralized 
assembly of point charges
is discussed in Appendix \ref{MadelungAppendix}.)
Note that $v^{corr}_0(r=0)$ is positive, 
reflecting the stabilizing contribution from the jellium compensation.
The value of $\nabla^2 v^{corr}_0(r=0)$ is then determined
from Eq. \ref{corrDifferential}:
\begin{equation}
\nabla^2 v^{corr}_0(r=0)=-\frac{4 \pi }{L^3}.
\end{equation}
Hence, the point-charge correction can be expanded as:
\begin{equation}
v^{corr}_0({\bf r})=\frac{\alpha_0 }{L}-\frac{2 \pi }{3 L^3}r^2+O(|{\bf r}|^4).
\label{PointChargeEquation}
\end{equation}
The terms in this parabolic expansion
bear a strong resemblance to those entering into the Makov-Payne correction \cite{MakovPayne1994}.
This correspondence is discussed further in Sec. \ref{EnergyCorrection}.


The above expansion allows us to approximate the electrostatic correction
induced by a set of compensating charges.
Indeed, introducing $N$ charges, we can define
a parabolic point-countercharge (PCC) 
potential $v^{corr}_{PCC}$ as:
\begin{equation}
v^{corr}_{PCC}({\bf r})=\sum_{n=1}^N q_n 
\left(\frac{\alpha_0}{L}-\frac{2 \pi }{3 L^3}({\bf r} -{\bf r}_n)^2\right).
\end{equation}
This expression may be rewritten:
\begin{equation}
v^{corr}_{PCC}({\bf r})=\frac{\alpha_0 q}{L} - \frac{2 \pi q}{3 L^3} r^2 + \frac{4\pi}{3L^3}{\bf p \cdot r} - \frac{2 \pi Q}{3 L^3},
\label{PCCExpansion}
\end{equation}
where $q=\sum_n q_n$ is the total charge, ${\bf p}=\sum_n q_n {\bf r}_n$ denotes 
the total dipole moment, and $Q=\sum_n q_n r_n^2$ 
stands for the total quadrupole moment of the countercharge distribution.
Eq. \ref{PCCExpansion} indicates that parabolic PCC schemes can
correct periodic-image errors up to quadrupole-moment order.
Note that no more than $N_{max}=7$ countercharges are sufficient
to obtain the most accurate parabolic correction (one charge for $q$,
two for ${\bf p}$, and four for $Q$). 
To obtain higher-order PCC corrections, one would need to
determine more terms in the expansion of the point-charge correction,
beyond the parabolic contributions. An example of
accurate calculations using harmonic expansions
can be found in Ref. \cite{CichockiFelderhof1989}. 


An alternative approach is to employ
countercharges whose corrective potential can be computed handily.
A popular choice is to use Gaussian densities, as proposed by
Bl\"ochl \cite{Blochl1995}.
Repeating the preceding analysis for a Gaussian density of charge $q=+e$, 
we can expand the Gaussian corrective potential $v^{corr}_{\sigma,L}$ as:
\begin{equation}
v^{corr}_{\sigma,L} = \frac{\alpha_{\sigma/L}}{L} - \frac{2 \pi }{3 L^3} r^2 +O(|{\bf r}|^4) ,
\end{equation}
where $\alpha_{\sigma/L}$ is the Madelung constant of an assembly of
Gaussians of width $\sigma$ immersed in a compensating jellium in
a cubic cell of length $L$.
It is more convenient, however, to write the corrective potential directly as:
\begin{eqnarray}
v^{corr}_{\sigma,L}({\bf r}) &=& v_\sigma({\bf r})-v'_{\sigma,L}({\bf r}) \nonumber \\
 &=& \frac{\textrm{erf}(r/\sigma)}{r} 
- \frac{1}{L^3}\sum_{\bf g \neq \bf 0} \frac{4\pi}{g^2}e^{-\sigma^2 g^2/4}e^{i{\bf g \cdot r}},
\end{eqnarray}
where $v_{\sigma}$ is the electrostatic 
potential of an isolated Gaussian charge, and
$v'_{\sigma,L}$ is the potential corresponding to a periodically repeated Gaussian
in a jellium background. 
The sum in the right-hand side of the equation converges very 
rapidly, and can be calculated
using FFT techniques. 
Superimposing $N$ compensating charges, 
the Gaussian-countercharge (GCC) corrective potential 
$v^{corr}_{GCC}$ can be expressed as:
\begin{equation}
v^{corr}_{GCC}({\bf r})=\sum_{n=1}^N q_n v^{corr}_{\sigma,L}({\bf r}-{\bf r}_n).
\end{equation}
This results in the following approximation for the open-boundary potential $v$:
\begin{equation}
v({\bf r}) \approx v'({\bf r}) + v^{corr}_{GCC}({\bf r}).
\label{GCCApproximation}
\end{equation}
We underscore that this scheme is equivalent to the Gaussian
scheme introduced by Bl{\"o}chl \cite{Blochl1995} and the LMCC
method proposed by Schultz \cite{Schultz1999}.
The equivalence with LMCC approach can be established 
by recasting Eq. \ref{GCCApproximation} as:
\begin{equation}
\left\{
\begin{array}{c}
v({\bf r}) \approx  v_{PBC}({\bf r}) + v_{GCC}({\bf r}) \\
\\
v_{PBC}({\bf r}) = v'({\bf r}) - v'_{GCC}({\bf r}),
\end{array}
\right.
\end{equation}
where $v_{GCC}({\bf r})=\sum q_n v_\sigma({\bf r}-{\bf r}_n)$ 
is the electrostatic potential generated by the isolated countercharge
distribution, and 
$v'_{GCC}({\bf r})=\sum q_n v'_{\sigma,L}({\bf r}-{\bf r}_n)$ is the corresponding
periodic potential. 


We are now in a position to compare the corrective
potentials $v^{corr}_{PCC}$ and $v^{corr}_{GCC}$ with the potential $v^{corr}$,
obtained as the direct difference between
the open-boundary potential and its periodic counterpart. For our
comparative analysis, we refer to the exact corrective potential $v^{corr}$ as the 
density-countercharge (DCC) potential. 
The DCC potential is obtained by evaluating the Coulomb
integral defining $v$ at each grid point in the unit cell. (A cheaper alternative
to this procedure is presented in the next section.)
The PCC, GCC, and DCC potentials for a charged pyridazine cation in a cubic cell
of length $L=15$ bohr are plotted in Figure \ref{PyridazineCationComparison}.
The PCC and GCC corrections are computed up to dipole order.
First, it should be noted that the
maximal energy of the PCC potential is slightly above its GCC counterpart, 
reflecting the fact that the Madelung
energy of an array of point charges immersed in a jellium is
higher than that of a jellium-neutralized array of Gaussian charges (cf.
Appendix \ref{MadelungAppendix}).
In addition, the maximal DCC
energy is found to be approximately 0.05 Ry above $\alpha_0 q / L$,
indicating that the dipole PCC and GCC corrections tend to underestimate
the energy of the system. Moreover,
the parabolic PCC potential is not as steep as its GCC counterpart, suggesting
that the energy underestimation will be more significant for 
the GCC correction. 
Owing to the cubic symmetry of the cell, 
the PCC and GCC potentials display the same curvature in
each direction of space, equal to one third of $-4\pi \avg{\rho}$.
In contrast, the curvature of the DCC potential 
is not uniform, due to the nonspherical nature of the molecular charge
density. This shape dependence suggests that the accuracy of
the GCC correction could be improved by optimizing the
geometry of the Gaussian countercharges.


In summary, we have shown that the PCC (Makov-Payne), GCC (LMCC), and DCC corrections
belong to the same class of periodic-image corrections. 
The analysis of the corrective potential has established that the parabolic PCC correction
cannot eliminate periodic-image interactions beyond quadrupole order.
Difficulties inherent in the GCC scheme have also been evidenced. 
To overcome these limitations, an efficient implementation of
the DCC correction is presented in Sec. \ref{DCCSection}. 

\subsection{Energy Correction}

\label{EnergyCorrection}

\begin{figure}
\includegraphics{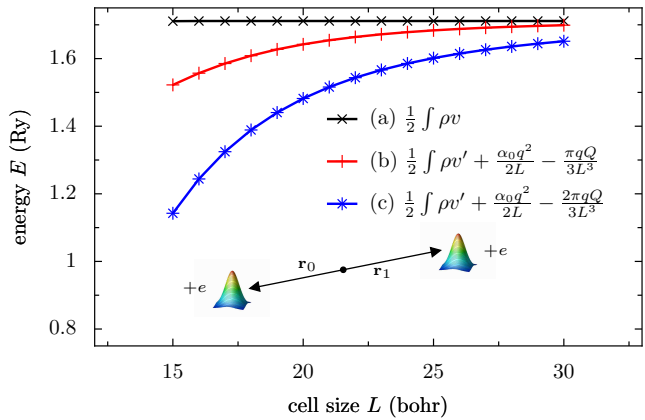}
\caption{Electrostatic energy of two Gaussians of unit charge
and unit spread calculated via (a) real-space
integration, (b) reciprocal-space integration
with the PCC energy correction
given by Eq. \ref{PCCCorrectionEquation}, and (c) reciprocal-space integration 
with the energy correction given by Eq. 15 in Ref. \cite{MakovPayne1994}.
The Gaussian charges are positioned at ${\bf r}_0=(-5,-5,-5)$
and ${\bf r}_1=(5,5,5)$ (corresponding to a quadrupole moment $Q$ of 153 
a.u.).
}
\label{MPTest}
\end{figure}


To conclude this preliminary analysis, we give the expression
of the energy correction $\Delta E^{corr}$ in terms of the corrective potential
$v^{corr}$.
The total electrostatic energy of the system being equal to:
\begin{equation}
E = \frac{1}{2} \int v({\bf r}) \rho({\bf r}) d{\bf r},
\end{equation}
the corrective energy can be expressed as \cite{Bengtsson1999}:
\begin{equation}
\Delta E^{corr} = \frac{1}{2} \int v^{corr}({\bf r}) \rho({\bf r}) d{\bf r}.
\label{EnergyCorrectionEquation}
\end{equation}


it is worth mentioning that in the case of a single 
point countercharge $q=\int \rho({\bf r}) d{\bf r}$, 
the PCC energy correction can be written as:
\begin{eqnarray}
\Delta E^{corr}_0 &=& \frac{1}{2} \int q v^{corr}_0({\bf r}) \rho({\bf r}) d{\bf r}
\nonumber \\
&=&  \frac{\alpha_0 q^2}{2L} - \frac{\pi qQ}{3 L^3}.
\label{PCCCorrectionEquation}
\end{eqnarray}
The first term corresponds to the Madelung energy correction, as proposed by Leslie
and Gillian \cite{LeslieGillian1985}. Note that the second term differs  from
Eq. 15 in Ref. \cite{MakovPayne1994} by a factor 1/2. The 
validity of the energy correction given by Eq. \ref{PCCCorrectionEquation}
is illustrated in Figure \ref{MPTest}.

\section{Implementation of the Density-countercharge Correction}

\label{DCCSection}

\subsection{Density-countercharge Algorithm}

\begin{figure}
\includegraphics[width=8.5cm]{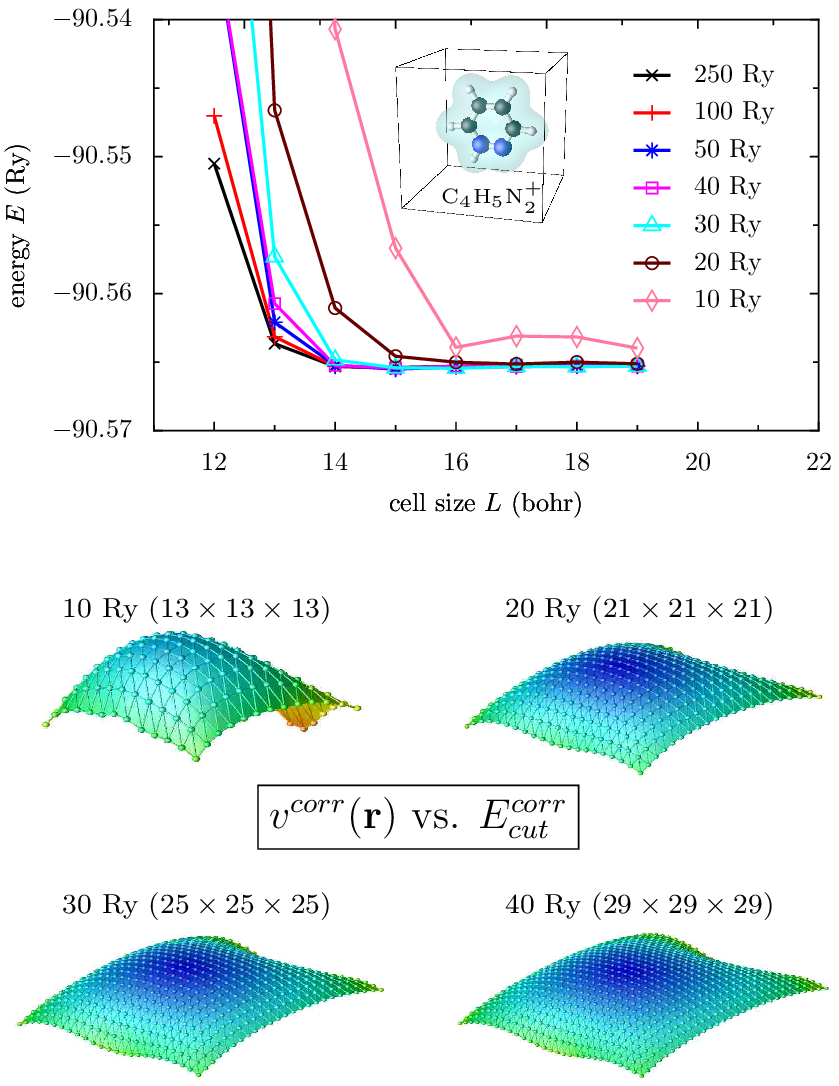} 
\caption{DCC total energy as a function of cell size for
a pyridazine cation varying the coarse-grid cutoff $E_{cut}^{corr}$ 
from 10 Ry ($M=13\times 13 \times 13$) to 250 Ry ($M=N=73\times 73 \times 73$).
Also depicted is the corrective potential $v^{corr}$ in the plane of the molecule
as a function of the coarse-grid resolution at a cell size of 15 bohr.}
\label{PyridazineCationEcorr}
\end{figure}


In the preceding section,
the corrective potential $v^{corr}=v^{corr}_{DCC}$ was calculated directly
by subtracting the periodic potential from its open-boundary counterpart. 
The computational cost of this direct method is prohibitively high,
on the order $O(N^2)$
(where $N$ is the number of grid points),
corresponding to the evaluation of Coulomb integrals
at each point of the grid. 
In this section, we present a scheme that
reduces this computational burden. 
The scheme exploits both the Poisson equation 
for $v^{corr}$ (Eq. \ref{corrDifferential})
and the fact that $v^{corr}$ is smoothly varying.


First, we note that taking into account appropriate boundary conditions, 
Eq. \ref{corrDifferential} can be solved efficiently
using multigrid solvers \cite{HolstSaied1993,HolstSaied1995,FattebertGygi2003,
ScherlisFattebert2006,BriggsHenson2000,TrottenbergOosterlee2001}.
Multigrid algorithms typically scale as
$O(N \log N)$, that is, comparable to the scaling of an
FFT computation. Hence,
the overall cost of the calculation can be reduced from $O(N^2)$ to $O(N^{5/3})$, 
corresponding to the expense arising from the 
determination of the boundary conditions.
Although a similar approach may be employed to directly solve the 
electrostatic equation defining $v$ (Eq. \ref{CoulDifferential}), we emphasize that
Eq. \ref{corrDifferential} allows a considerable reduction in numerical error
in the finite-difference evaluation of the electronic Laplacian---since 
$v^{corr}$ is much smoother than $v$.


Further exploiting this idea, it is possible to solve
Eq. \ref{corrDifferential} on a grid much coarser than that 
used to discretize the charge density. 
To illustrate this fact, we consider a pyridazine cation in a periodic
cubic cell of varied size (Figure \ref{PyridazineCationEcorr}).
The total energy of the system is calculated
using density-functional theory \cite{PayneTeter1992}. 
An energy cutoff $E_{cut}=250$ Ry is applied 
to the plane-wave expansion of the charge density. The total energies are corrected
using the DCC scheme by solving the electrostatic equation of $v^{corr}$
on a coarse grid for several values of the energy cutoff, denoted $E^{corr}_{cut}$.
Reducing the energy cutoff $E^{corr}_{cut}$ from 250
to 40 Ry, the corrected energies are observed to
depart by less than $5\times10^{-3}$ Ry from their converged values for
cell sizes greater than 13 bohr. 
The ability to decrease
the number of grid points without a significant loss of accuracy enables a substantial
reduction of the additional computational cost from $O(N^{5/3})$ to 
$O(M^{5/3})$, where $M$ is the number of coarse-grid points. Note that 
diminishing the plane-wave energy cutoff from 250 to 40 Ry at $L=15$ bohr
reduces the cost of the boundary-condition calculation
by a factor $29^5/73^{5} \approx 1/100$.


Before presenting the algorithm,
we draw attention to the fact that the DCC scheme relies
on the central idea that most of the structural characteristics
of the open-boundary potential $v$ can be removed by subtracting out 
its periodic counterpart $v'$.
The residual $v^{corr}$
(that is, the amount by which $v'$ fails to reproduce $v$) 
is smooth and can be determined
on a coarse grid at low computational cost. 
Additional computational savings come from the ability
to avoid updating the potential $v^{corr}$
at each step of the self-consistent-field (SCF)
calculation, but instead at fixed interval between electronic iterations.


The DCC algorithm for a typical
electronic-structure
calculation can be described as follows.
Let $N^{corr}$ denote the number of SCF steps
between each update of the corrective potential. 
\begin{enumerate}
\item Start from an initial charge distribution $\rho$
on the fine grid.
\item
Calculate the periodic potential $v'$ corresponding to $\rho$.
\item Transfer $\rho$ and $v'$ on the coarse grid (tricubic interpolation 
\cite{NumericalRecipes})
to obtain the coarse-grid density $\tilde \rho$ and coarse-grid periodic potential $\tilde v'$.
\item Calculate the real-space potential $\tilde v$
at the boundaries of the coarse grid from $\tilde \rho$
to obtain the Dirichlet boundary conditions
$\tilde v^{corr}=\tilde v-\tilde v'$.
\item Solve $\nabla^2 \tilde v^{corr}= - 4 \pi \langle \rho \rangle$ (multigrid techniques)
to obtain the corrective potential $\tilde v^{corr}$.
\item Transfer $\tilde v^{corr}$ on the fine grid (tricubic interpolation)
to obtain $v^{corr}$, and calculate 
$v=v^{corr}+v'$.
\item Perform $N^{corr}$ electronic SCF steps.
\item Iterate from Step 2 until reaching SCF convergence.
\end{enumerate}
Note that we employ real-space tricubic interpolation techniques
in order to avoid oscillatory distortions inherent in Fourier-transform interpolation schemes.
We also underscore that the DCC algorithm can be efficiently parallelized, 
since its most expensive step (namely, the calculation
of the Dirichlet boundary conditions)
scales linearly with the number of processors.


The above procedure can be adapted to 
one- and two-dimensional systems by considering the linear or
planar average of the charge density for calculating the corrective potential
\cite{BaldereschiBaroni1988}.
(The validity
the linear- or planar-average approximations will be discussed in the final section.)
The computational cost of this approach is moderate, on the order of
$O(M^{1/3})$ and $O(M)$ for one and two dimensions, respectively. 


It should also be mentioned
that the DCC algorithm can be
used in combination with multipole-expansion
methods
for a rapid evaluation of the Dirichlet boundary conditions (Step 4). The accuracy of
this approach depends on the precision of
the multipole expansion at the boundary
of the supercell.
(A mathematical discussion on the
long-range accuracy of multipole expansions 
is presented in Sec. 3.4. of Greengard's dissertation
\cite{Greengard1988}.)
The performance the multipole-expansion approach is reported
in Appendix \ref{MultipoleAppendix}.

\subsection{Applications}

\label{DCCApplicationSection}

\begin{figure}
\includegraphics[width=8.5cm]{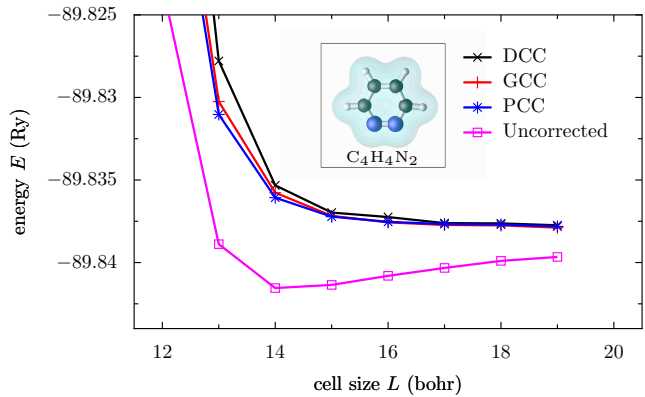}
\caption{
Total energy as a function of cell size for 
a neutral pyridazine molecule without correction and corrected using the PCC, GCC, and
DCC schemes.
The PCC and GCC corrections
are calculated up to quadrupole order.
The inset
shows a pyridazine molecule in a cell of size $L=15$ bohr.
}
\label{PyridazineConvergence}
\end{figure}

\begin{figure}
\includegraphics[width=8.5cm]{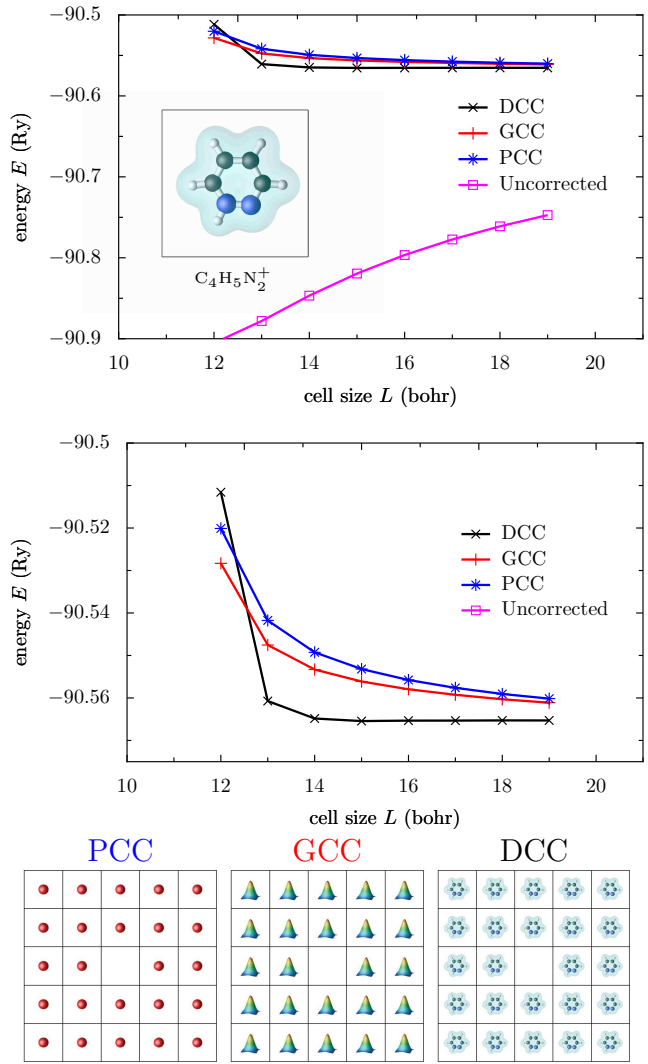}
\caption{
Total energy as a function of cell size for
a pyridazine cation without correction and corrected using the PCC, GCC, and
DCC schemes.
The PCC and GCC corrections
are calculated up to quadrupole order.
The inset of the top graph shows a pyridazine cation in a cell of size $L=15$ bohr.
}
\label{PyridazineCationConvergence}
\end{figure}

\begin{figure}
\includegraphics[width=8.5cm]{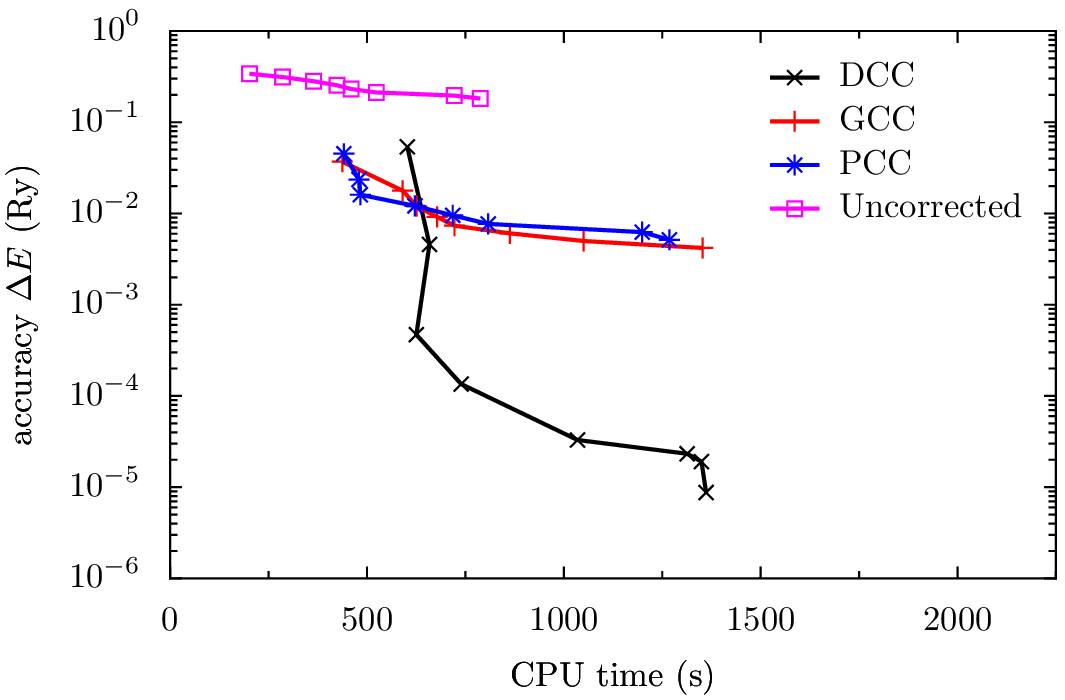} 
\caption{Accuracy of the total energy as a function
of computational time without correction and
using the PCC, GCC, and DCC schemes for cell sizes in the range
12-19 bohr.
For each scheme, the corrective potential
is updated every five SCF iterations.}
\label{PyridazineCationLog}
\end{figure}


The energy of a pyridazine molecule as a function of cell size $L$ 
for each countercharge correction is reported in Figure \ref{PyridazineConvergence}. 
For this neutral species, the uncorrected energy shows a characteristic minimum
at $L=$ 14 bohr before slowly approaching its asymptotic value.
In contrast, the corrected energies are seen
to converge monotonically towards their common energy limit. Although the three
schemes demonstrate comparable convergence, it should be noted that the PCC method is 
slightly more accurate. In addition to 
further validating the energy expansion given by Eq. \ref{PCCCorrectionEquation}, 
this comparison suggests that the PCC correction can be preferred for 
studying neutral species, with the notable exception of
elongated systems ({\it e.g.}, polymer fragments or terminated nanotubes).


We now consider the energy of a pyridazine cation
as a function of cell size (Figure \ref{PyridazineCationConvergence}).
We use energy cutoffs of 35 and 250 Ry for expanding the wavefunctions
and the charge density, and select a coarse-grid cutoff of 35 Ry for calculating
the DCC correction.
Expectedly, the uncorrected energy converges very slowly with respect to $L$ (at 19 bohr,
the energy error is still larger than 0.15 Ry). The PCC
and GCC corrections substantially improve the convergence of the total energy,
reducing periodic-image errors by one order of magnitude.
Using the DCC scheme, the energy is 
observed to converge even more rapidly, reflecting the exponential disappearance
of energy errors arising from charge density spilling across periodic cells:
at a cell size of 15 bohr, which is barely larger than the
size of the molecule, the DCC energy is converged within $10^{-4}$ Ry. 
The performance of each scheme 
as a function of the total computational time 
is shown on a logarithmic energy scale in Figure \ref{PyridazineCationLog}.
Each curve corresponds to cell sizes in the range 
12-19 bohr. For meaningful comparison with the DCC scheme, the PCC and GCC
corrective potentials are also updated at fixed SCF intervals.
We observe that the computational cost of the corrected calculations is 
comparable to that without correction
for a considerable improvement in accuracy.
For this charged system, 
the DCC approach constitutes the most advantageous alternative, improving the
energy precision by two orders of magnitude over the PCC and GCC corrections
for cell sizes above 15 bohr. 
 

\begin{figure}
\includegraphics[width=8.5cm]{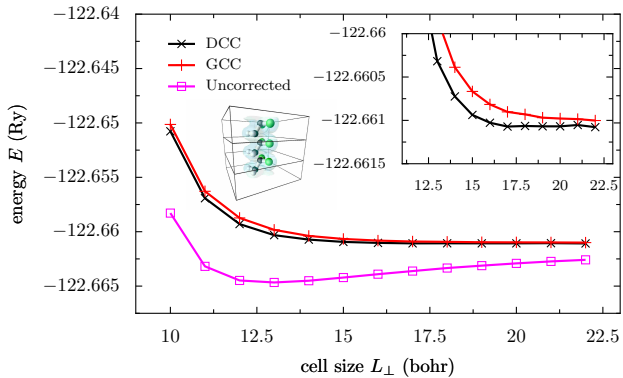}
\caption{Total energy as a function of transverse cell size for
a polyvinylidene fluoride (PVDF) chain without correction,
and using the GCC and DCC schemes.}
\label{PVDFEnergy}
\end{figure}

The performance of the DCC and GCC corrective schemes for
a neutral polyvinylidene fluoride (PVDF) chain
is reported in Figure \ref{PVDFEnergy}. The comparison 
shows a significant improvement of energy convergence for
both schemes. As shown in the inset,
the performance of the DCC scheme is perceptibly superior to
that of the GCC scheme. We emphasize that for systems
exhibiting one dimensional periodicity,
the additional computational cost due to the electrostatic
correction is moderate,
on the order of $O(M)$ at most.


\begin{figure}
\includegraphics{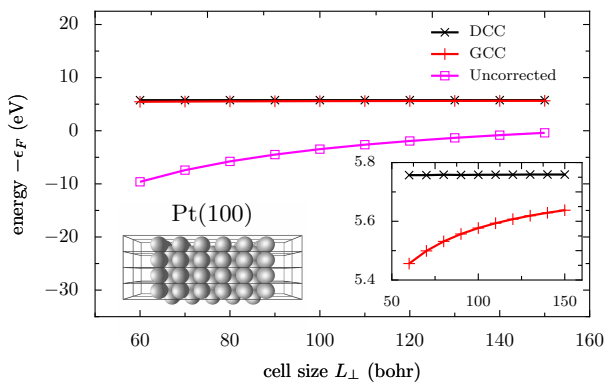}
\caption{Convergence of the opposite Fermi energy $-\epsilon_F$
as a function of transverse cell size 
for a Pt(100) slab  without correction,
and using the GCC and DCC schemes.}
\label{Pt100WorkFunction}
\end{figure}

The DCC scheme can also
be used in the calculation of work functions, as it solves energy-reference issues
by automatically setting the vacuum
level to zero. Figure \ref{Pt100WorkFunction} depicts 
the convergence of the opposite Fermi energy of a Pt(100) slab as 
a function of transverse cell size. The wavefunction, charge-density,
and corrective potential energy cutoffs are 25, 200,
and 150 Ry, respectively. We use a shifted $5 \times 5 \times 1$
mesh with a cold-smearing occupation function \cite{Marzari1996}
(smearing temperature of 0.03 Ry) to sample the 
Brillouin zone.
Without correction, the relative error
in the Fermi energy stays above 100\% for all cell sizes in
the considered range. Using
the GCC scheme, the convergence of the Fermi level
improves greatly: at 150 bohr, the relative error
reduces to approximately 0.1 eV. Employing the DCC 
corrective scheme, the calculated Fermi energy is converged within 2 meV at 60 bohr
and 0.1 meV at 150 bohr. Thus,
the DCC scheme allows to directly determine the
work function of a metal as the opposite of the calculated Fermi energy 
using supercells of minimal size.
A similar convergence improvement
is obtained for the work function of carbon nanotubes \cite{SinghMillerMarzari2007}.
Besides improving the convergence of total energies,
the DCC approach
can be employed to correct structural and vibrational properties \cite{RozziVarsano2006},
and to calculate linear-response characteristics with a reduced computational effort
\cite{RozziVarsano2006,IsmailBeigi2006,Kozinskymarzari2006}. 

\section{Beyond the Linear- and Planar-average Approximations}

\label{AverageApproximation}

\subsection{Treating Systems with Partial Periodicity}

In the preceding sections, we have assumed that the corrective
potential of a one- or two-dimensional system 
can be obtained by
homogenizing the system along its periodicity directions,
as initially proposed by Baldereschi, Baroni, and Resta \cite{BaldereschiBaroni1988}. 
This approach, referred to as the linear-
or planar-average approximation, has been frequently employed
in electronic-structure calculations
\cite{BaldereschiBaroni1988,PeressiBaroni1990,NeugebauerScheffler1992,LozovoiAlavi2003,Bengtsson1999}.

Alternative schemes adapting the Ewald method to
evaluate conditionally convergent lattice sums
 \cite{JaffeHess1996} 
or generalizing the FMM approach \cite{KudinScuseria1999,KudinScuseria2004}
have also been proposed for systems exhibiting partial periodicity.
Such schemes are particularly suited to
localized-orbital calculations but
are of relatively limited applicability for plane-wave implementations.
Here, we propose an efficient method to calculate the 
electrostatic potential for partially periodic systems,
taking into account the full three-dimensional structure
of the charge distribution. In addition to presenting this methodological
extension, we discuss how to assess the validity of the
linear- and planar-average approximations {\it a priori} in terms of
structural characteristics of the system.

\subsection{DCC Scheme for One-dimensional Periodicity}

\begin{figure*}
\includegraphics[height=13cm]{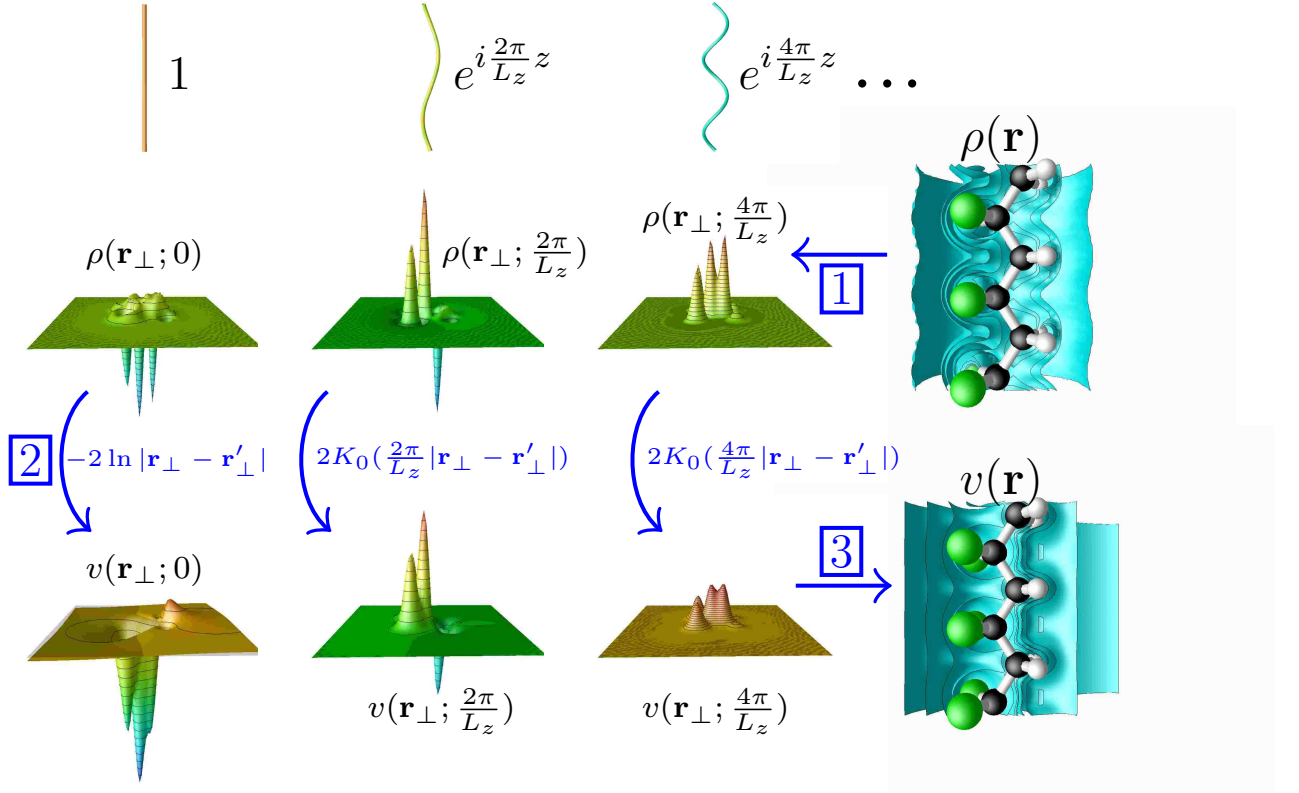}
\caption{Fourier-decomposition calculation of the electrostatic potential 
$v({\bf r}_\perp,z)=\sum_{g_z} v({\bf r}_\perp;g_z)
e^{ig_zz}$ for an infinite polyvinylidene fluoride (PVDF) chain.
(1) The longitudinal Fourier transform of the charge density
is calculated to obtain the contributions from each axial
wavevector $g_z$; (2) the electrostatic potential generated by
each Fourier component of the charge density is calculated 
using Green's functions; (3) the electrostatic potential
is then transformed back to real space.
\label{BesselMethod}}
\end{figure*}


To introduce
the DCC approach for one-dimensional systems, we first study
the electrostatic problem corresponding to an isolated sinusoidal-density line:
\begin{equation}
\rho({\bf r})=\delta^{(2)}({\bf r_\perp})\exp(ig_zz),
\end{equation}
where $\delta^{(2)}$ stands for the two-dimensional Dirac delta function and
${\bf r_\perp}$ denotes the transverse coordinates $(x,y)$. 
Making the ansatz $v({\bf r})={\cal G}({\bf r_\perp};g_z)\exp(ig_zz)$ 
for the Green's function, we obtain:
\begin{equation}
( \nabla^2_\perp - g_z^2 ) {\cal G}({\bf r_\perp};g_z) 
= - 4 \pi \delta^{(2)}({\bf r_\perp}).
\label{GeneralizedPoisson}
\end{equation}
The solution of this generalized electrostatic problem can be written as:
\begin{equation}
\left\{
\begin{array}{cccl}
{\cal G}({\bf r}_\perp;0)&=&-2 \ln|{\bf r}_\perp|, & \\ 
\\
{\cal G}({\bf r}_\perp;g_z)&=&2 K_0(g_z |{\bf r}_\perp|)
& \textrm{for $g_z \neq 0$}
\label{GreenFunctions}
\end{array}
\right.
\end{equation}
where $K_0$ is the modified Bessel function of the second kind.
Note that $K_0(g_z |{\bf r}_\perp|) = - \ln|{\bf r}_\perp| + ...$ 
when $g_z |{\bf r}_\perp|$ approaches zero, reflecting the fact that a sinusoidal-density 
line can be considered as uniform when seen from a distance much 
smaller than its wavelength. 
Knowing the electrostatic potential generated by a single line
(the Green's function characterizing the generalized electrostatic problem),
the potential of an arbitrary one-dimensional 
charge distribution can be determined analytically,
as illustrated in Figure \ref{BesselMethod}. 
The general procedure consists of calculating the 
one-dimensional Fourier transform of $\rho$
to obtain its longitudinal Fourier components $\rho({\bf r}_\perp;g_z)$ (step 1).
Each individual components is then
convoluted with the electrostatic potential generated by a sinusoidal density, 
as expressed in Eq. (\ref{GreenFunctions})
to obtain the Fourier components $v({\bf r}_\perp;g_z)$
of the open-boundary potential (step 2):
\begin{widetext}
\begin{equation}
\left\{
\begin{array}{cccl}
v({\bf r}_\perp;0)&=& -2 \displaystyle\int
\ln| {\bf r}_\perp - {\bf r}_\perp'| \rho({\bf r}_\perp';0)
d{\bf r}_\perp', & \\
v({\bf r}_\perp;g_z)&=& 2 \displaystyle\int 
K_0(g_z | {\bf r}_\perp - {\bf r}_\perp'|) 
\rho({\bf r}_\perp';g_z) d{\bf r}_\perp'
& \textrm{for $g_z \neq 0$}.
\end{array}
\right.
\label{Convolution2D}
\end{equation}
\end{widetext}
Finally, the open-boundary potential is transformed back to real space (step 3).
We underscore that this procedure directly extends the linear-average approximation
since the linear average of the charge density corresponds to
the first term of the one-dimensional Fourier decomposition. 
Thus, averaging the charge density along
the axis of periodicity amounts to restricting
the Fourier series to its $g_z=0$ term. 


To estimate errors resulting from this truncation,
we analyze the asymptotic behavior of $v({\bf r}_\perp;g_z \neq 0)$
at large $g_z|{\bf r}_\perp|$:
\begin{equation}
v({\bf r}_\perp;g_z) \approx
\sqrt{\frac{\pi}{2}} \frac{e^{-g_z|{\bf r}_\perp|}}{\sqrt{g_z|{\bf r}_\perp|}}
 \textrm{ when $g_z|{\bf r}_\perp| \gg 1$}.
\label{AsymptoticBessel}
\end{equation}
From Eq. \ref{AsymptoticBessel}, the validity of the linear average approach
can be assessed by calculating the ratio of the cell size in the transverse
direction $L_\perp$ (that is, the distance between
periodic replicas) to the typical wavelength $\lambda_\parallel$
characterizing longitudinal inhomogeneities in the system.
For large values of the dimensionless parameter 
$L_\perp/\lambda_\parallel$, periodic-image interactions
are predominantly due to the logarithmic first-order contribution 
$v({\bf r}_\perp;0)$ corresponding to the linear average of the
charge density. Thus, as expected intuitively, the linear-average 
approximation is valid in this situation. In contrast, 
when $\lambda_\parallel$ is comparable to the distance $L_\perp$
between periodic images, higher-order Fourier components
$v({\bf r}_\perp;g_z)$ corresponding to
$g_z \approx 2\pi/\lambda_\parallel$ must also be taken into 
consideration. 


Despite its merit in discussing the validity of
the linear-average approximation, 
determining the open-boundary potential using the preceding approach requires
expensive summations for each point ${\bf r}_\perp$
of the two-dimensional grid and for each longitudinal wavevector $g_z$.
Along the same methodological lines as those of the DCC algorithm, a substantial
reduction of computational cost can be achieved by 
exploiting the
periodic potential $v'$, whose longitudinal Fourier components
can be computed inexpensively using FFT techniques:
\begin{widetext}
\begin{equation}
\left\{
\begin{array}{cccl}
v'({\bf r}_\perp;0) & = & \displaystyle\sum_{{\bf g}_\perp \neq {\bf 0}} \frac{4\pi}{{\bf g}_\perp^2} 
\rho({\bf g}_\perp)e^{i{\bf g}_\perp \cdot {\bf r}_\perp}, & \\
v'({\bf r}_\perp;g_z)& = & \displaystyle\sum_{{\bf g}_\perp} \frac{4\pi}{{\bf g}_\perp^2 + g_z^2}
\rho({\bf g}_\perp + g_z \hat {\bf z} )e^{i{\bf g}_\perp \cdot {\bf r}_\perp} 
& \textrm{for $g_z \neq 0$}.
\end{array}
\right.
\end{equation}
\end{widetext}
After coarse-grid interpolation, the component of
the open-boundary potential $v({\bf r}_\perp;g_z)$
can be calculated at the boundaries of the domain, yielding 
Dirichlet boundary conditions for the smooth corrective components 
$v^{corr}({\bf r}_\perp;g_z)=v'({\bf r}_\perp;g_z)-v'({\bf r}_\perp;g_z)$. 
The corresponding $g_z$-dependent electrostatic problems read:
\begin{equation}
\left\{
\begin{array}{cccl}
\nabla^2 v^{corr}({\bf r}_\perp;0) & = & - 4 \pi \avg{\rho} & \\
\\
( \nabla^2 - g_z^2 ) v^{corr}({\bf r}_\perp;g_z) & = & 0 &
\textrm{for $g_z \neq 0$}
\end{array}
\right.
\end{equation}
These differential equations can be solved using efficient
multigrid techniques.
Once calculated, the longitudinal Fourier components of the 
electrostatic correction are added to those of the periodic
potential, thereby recovering $v({\bf r}_\perp;g_z)$.
Finally, the potential $v({\bf r})$ is computed via an
inverse Fourier transform. 

\subsection{DCC Scheme for Two-dimensional Periodicity}

The electrostatic potential of a slab can be calculated in 
real space using a scheme similar to that presented above.
The formalism is to a great extent analogous to that developed
by Lang and Kohn for studying  interactions between 
localized external charges and metallic surfaces \cite{LangKohn1973},
and to the Green's function approach recently 
proposed by Otani and Sugino \cite{OtaniSugino2006}.
The prescription consists of performing two-dimensional
Fourier transforms to obtain the charge-density profile $\rho(z;{\bf g}_\parallel)$ 
associated with each wavevector ${\bf g_\parallel}=(g_x,g_y)$ parallel
to the surface.
Solving the electrostatic problem for sinusoidal density layers,
the two-dimensional Green's functions ${\cal G}(z;{\bf g}_\parallel)$ 
can be written as:
\begin{equation}
\left\{
\begin{array}{cccl}
{\cal G}(z;{\bf 0})&=&\displaystyle - 2 \pi |z|, & \\
{\cal G}(z;{\bf g}_\parallel)&=&\displaystyle 2 \pi \frac{e^{-g_\parallel |z|}}{ g_\parallel}
& \textrm{for $g_\parallel \neq 0$}.
\end{array}
\right.
\label{GreenFunction1D}
\end{equation}
Hence, as in the one-dimensional case, the density-average approximation
is valid provided that the geometrical parameter $L_\perp/\lambda_\parallel$
is large---this criterion is identical to that derived by Natan, Kronik, and Shapira
\cite{NatanKronik2000}.
In addition, the above expressions allow one to determine the corrective potential
of a two-dimensional system by integrating the differential equations:
\begin{equation}
\left\{
\begin{array}{cccl}
\frac{d^2}{dz^2} v^{corr}(z;{\bf 0}) & = & - 4 \pi \avg{\rho} & \\
\\
( \frac{d^2}{dz^2} - g_\parallel^2 ) v^{corr}(z;{\bf g}_\parallel) & = & 0 &
\textrm{for $g_\parallel \neq 0$}
\end{array}
\right.
\label{TwoDimEquation}
\end{equation}
Parenthetically, it is important to note
that Eq. \ref{TwoDimEquation} can be solved analytically, taking into account
the boundary conditions calculated by superposition---that is,
by convoluting the longitudinal components of ${\cal G}$ 
and $\rho$ (similarly to Eq. \ref{Convolution2D}),
then subtracting out the components of $v'$. Therefore, the additional cost
of the two-dimensional DCC correction is negligible.

\subsection{Applications}

\begin{figure}
\includegraphics[width=8.5cm]{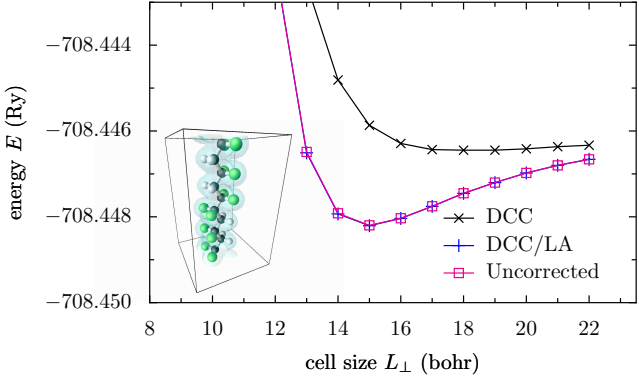}
\caption{Total energy as a function of transverse cell size for
a $\rm -[CH_2CF_2]_3-[CF_2CH_2]_3-$ polymer chain without correction,
corrected using the density-countercharge scheme with full Fourier decomposition (DCC), and 
by limiting the density-countercharge decomposition to the linear-average $\bf g= 0$ component (DCC/LA).}
\label{PVDFLongEnergy}
\end{figure}

\begin{figure}
\includegraphics[width=8.5cm]{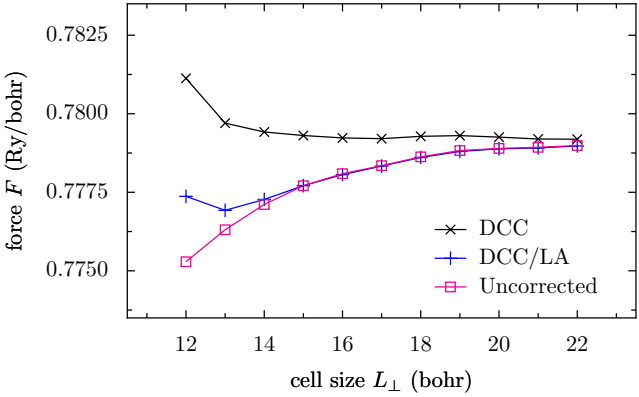}
\caption{Force on one of the fluorine atoms along 
a transverse lattice direction
as a function of transverse cell size for
a $\rm -[CH_2CF_2]_3-[CF_2CH_2]_3-$ polymer chain without correction,
corrected using the density-countercharge scheme with full Fourier decomposing (DCC), and
by limiting the density-countercharge decomposition to the linear-average $\bf g= 0$ component (DCC/LA).}
\label{PVDFLongForce}
\end{figure}


The convergence of the total energy with respect
to transverse cell size for a fluoropolymer chain $\rm -[CH_2CF_2]_3-[CF_2CH_2]_3-$
of long periodicity $\lambda_\parallel \approx 24$ bohr is depicted in Figure \ref{PVDFLongEnergy}. We employ ultrasoft pseudopotentials  \cite{Vanderbilt1990}
with energy cutoffs of 50 and 500 Ry for the plane-wave expansions
of the electronic wavefunctions and charge density, respectively. The energy cutoff 
for calculating the corrective potential is 80 Ry. We use
a shifted $1 \times 1 \times 2$ mesh with cold-smearing occupations
\cite{Marzari1996} (smearing temperature
of 0.02 Ry).
Within the linear average approximation (DCC/LA),
the corrected energy closely coincides with the uncorrected energy
due to the absence of polarization in the longitudinal average of the charge density.
For the cell parameters considered, the geometrical ratio $L_\perp/\lambda_\parallel$
varies from 0.5 to 0.9, that is, beyond 
the range of validity of the linear average approximation.
As a result, we observe that the DCC/LA energy
converges slowly towards its asymptotic value. In contrast, the DCC scheme
with full Fourier decomposition significantly improves the convergence of the total energy
(at 16 bohr, the accuracy of DCC energy is 
approximately $5 \times 10^{-5}$ Ry whereas that of the uncorrected and DCC/LA energies
is approximately $10^{-3}$ Ry). Figure \ref{PVDFLongForce} depicts the convergence
of the force on one of the fluorine atoms.
Similarly to the convergence of the total energy, the atomic-force convergence 
is seen to improve substantially by applying the
DCC correction: at 16 bohr, the DCC
force is converged within less than $10^{-4}$ Ry/bohr, while that obtained 
without correction or using the DCC/LA scheme are converged within $10^{-3}$ Ry/bohr.
We underscore that the additional computational cost of the DCC correction is moderate.
Indeed, at 16 bohr, the additional computational cost is $\sim$8\%.


\begin{figure}
\includegraphics[width=8.5cm]{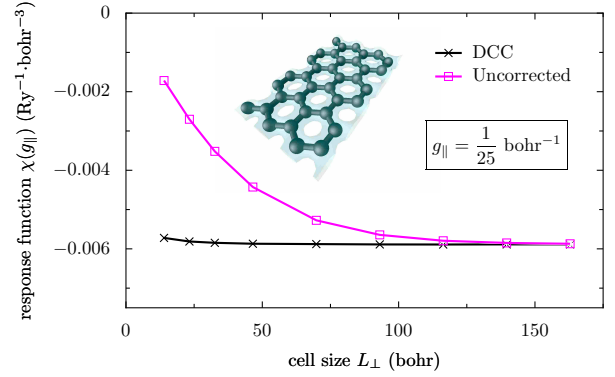}
\caption{Longitudinal density response coefficient 
$\chi(g_\parallel)={\partial n(g_\parallel)}/{\partial
v(g_\parallel)}$ as a function of transverse cell size for
a graphene sheet without correction,
and corrected using the density-countercharge scheme with full Fourier decomposition (DCC).}
\label{ResponseGraphene}
\end{figure}

To conclude this study, we consider
the electronic density response of a graphene sheet subject to a perturbation field.
Figure \ref{ResponseGraphene} reports the dependence of the linear-response coefficient 
$\chi(g_\parallel)={\partial n(g_\parallel)}/{\partial 
v(g_\parallel)}$ with respect to the interplane
distance $L_\perp$ for a longitudinal sinusoidal perturbation of wavevector 
${g_\parallel}=\frac{1}{25}$ 
bohr$^{-1}$. 
The wavelength of the perturbation field being large ($\lambda_\parallel=157$ bohr),
the uncorrected response coefficient does not convergence until reaching cell
sizes on the order of hundreds of bohrs. Contrary to uncorrected calculations, 
the DCC-corrected linear response shows considerable 
convergence improvement with a negligible increase in computation cost. 
For comparison, at an interplane distance of $L_\perp=50$ bohr, 
the relative error in the uncorrected linear-response coefficient 
$\chi(g_\parallel)$ is on the order of 
25\%, while it is lower than $1\%$ using the DCC correction. 

\section{Conclusion}


We have studied the analytical properties of the corrective potential,
defined as the difference between the electrostatic potential and
its periodic counterpart, unifying the Makov-Payne (PCC) and LMCC 
(GCC) schemes in
the same class of periodic-image corrections and suggesting possible improvements
for both methods.
Based on these properties, we have shown that the 
periodic-image errors can be eliminated 
at a moderate computational cost of $O(M^{5/3})$,
where $M$ is the number of points of the mesh used in the calculation the corrective potential, which is
generally about two orders of magnitude
smaller than the number of points of the charge-density grid.
The resulting density-countercharge (DCC) scheme 
owes its improved efficiency to the determination of the
exact boundary conditions characterizing the electrostatic potential.
In several cases of interest, we have shown
that the DCC algorithm represents a
beneficial compromise between cost and accuracy.
The validity of the linear- and planar-average approximations
routinely employed in the study of partially periodic systems 
has also been discussed. An efficient 
scheme going beyond these conventional approximations for inhomogeneous systems
has been proposed and validated.


Relevant applications for the DCC algorithm
include the study of molecular adsorption at solid-vacuum interfaces
in the constant-charge regime, the determination of structural
parameters, the correction of vibrational spectra,
the inexpensive calculation of work functions, 
and the determination of linear-response properties with a
reduced computational effort.

\begin{acknowledgments}
The calculations in this work have been performed using the 
Quantum-Espresso package \cite{Espresso} (GNU General Public License),
and the conjugate-gradient multigrid solver developed by M. Holst,
as part of the Parallel algebraic Multigrid/Finite-element Toolkit 
\cite{FeTK}. Both software packages are licensed for use under
the GNU General Public License.

The authors acknowledge support from the MURI grant DAAD 19-03-1-0169, 
NSF-NIRT DMR-0304019, and ISN-ARO grant DAAD 19-02-D-0002. 
I. D. personally thanks the \'Ecole Nationale des Ponts et Chauss\'ees (France) 
and the Martin Family Society of Fellows for Sustainability for their help and support. 
Comments and suggestions from Jean-Luc Fattebert and \'Eric Canc\`es
about the use of multigrid algorithms,
and from Raffaele Resta about electrostatics in periodic boundary conditions 
are gratefully acknowledged.
The authors thank Brandon Wood, Nicolas Poilvert, Young-Su Lee, Arash Mostofi, Oswaldo Dieguez,
and Damian Scherlis for valuable comments and suggestions.
\end{acknowledgments}

\appendix

\section{Madelung constants and Gaussian potentials}

\label{MadelungAppendix}


In this appendix, we determine the Madelung constants
of periodic point charges immersed in
a compensating jellium background in one, two, and three dimensions
for lattices characterized by a single geometric parameter $L$. 
A compilation of high-precision values for these fundamental
constants is generally not found in the literature.


These values are computed using the asymptotic expansion of the 
Madelung constant $\alpha_{\sigma/L}$ of an array of
Gaussian charges of spread $\sigma$ in a compensating jellium,
which is defined as:
\begin{equation}
\alpha_{\sigma/L} = (v_{\sigma}(0) - v'_{\sigma,L}(0))L^{d-2},
\end{equation}
where $d$ is the spatial dimension.
To obtain the expansion of $\alpha_{\sigma/L}$ in the limit $\sigma/L \rightarrow 0$,
we may write $v'_{\sigma,L}(0)$ as:
\begin{equation}
v'_{\sigma,L}(0)=\frac{L^{2-d}}{\Omega_d} w_d(\frac{\sigma^2}{L^2}),
\label{ZeroPotential}
\end{equation}
\begin{equation}
w_d(\frac{\sigma^2}{L^2}) = \sum_{{\bf g' }\neq{\bf 0}} \frac{4\pi}{g'^2}\exp(-\frac{g'^2}{4}  \cdot \frac{\sigma^2}{L^2}),
\end{equation}
where $\Omega_d$ is the volume of $d$-dimensional
unit cell, and ${\bf g'}=L {\bf g}$ denotes the dimensionless wavevector.
Differentiating $w_d$ with respect to $\sigma^2/L^2$, we obtain:
\begin{eqnarray}
\frac{dw_d}{d(\sigma^2/L^2)}&=&- \pi \sum_{{\bf g' }\neq{\bf 0}} \exp(-\frac{g'^2}{4} \cdot \frac{\sigma^2}{L^2}) \nonumber \\
& = & \pi - \pi \sum_{\bf g' } \exp(-\frac{g'^2}{4} \cdot \frac{\sigma^2}{L^2}).
\end{eqnarray}
In the limit $\sigma/L \rightarrow 0$, this derivative becomes:
\begin{equation}
\frac{dw_d}{d(\sigma^2/L^2)}=\pi - \frac{\Omega_d}{\pi^{d-1}}\left(\frac{\sigma^2}{L^2}\right)^{-d/2}
\int_{R^d} e^{-u^2} d{\bf u}+...
\end{equation}
Integrating this expression, 
we obtain the asymptotic expansions of
$v'_{\sigma,L}(0)$ and $\alpha_{\sigma/L}$ listed in Table \ref{MadelungTable}.


\begin{figure}
\includegraphics[width=8.5cm]{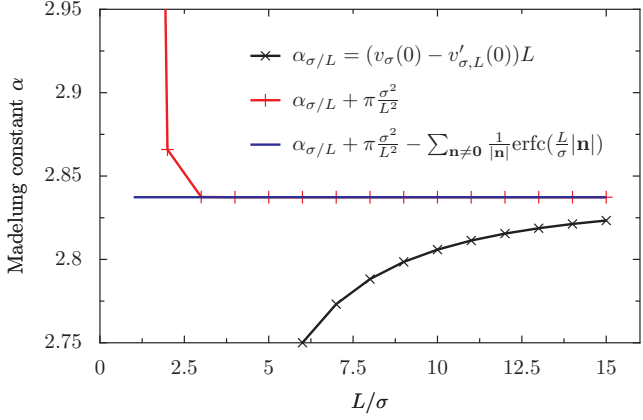}
\caption{Convergence of the Madelung constant as a function
of the geometric parameter $L/\sigma$ for a cubic
unit cell using the approximation given by Eq. \ref{MadelungApproximant}.
(The black curve is Eq. \ref{MadelungApproximant} without the $\pi \sigma^2/L^2$ and 
the complementary-error-function terms, the red curve is 
Eq. \ref{MadelungApproximant} without the complementary-error-function term, 
and the blue curve is Eq. \ref{MadelungApproximant}).
Note the negligible contribution of the complementary-error-function
term beyond $L/\sigma=3$ and the improvement in convergence brought about
by the term $\pi \sigma^2/L^2$.}
\label{MadelungCalculation}
\end{figure}

Hence, the Madelung constant $\alpha_0$ can be calculated with high accuracy
from the expansion of $\alpha_{\sigma/L}$. In the case of a cubic lattice
of point charges, we obtain:
\begin{eqnarray}
\alpha_0 & \approx & \alpha_{\sigma/L} + \frac{\pi \sigma^2}{L^2}
- \sum_{{\bf n}\neq {\bf 0}}
\frac{1}{|{\bf n}|}\textrm{erfc}(\frac{L}{\sigma}|{\bf n}|) \nonumber \\
& \approx & \frac{1}{L^2}\sum_{{\bf g}\neq {\bf 0}}
\frac{4\pi}{g^2}e^{-\sigma^2g^2/4} - \frac{2L}{\sqrt{\pi}\sigma} \nonumber \\
&&+ \frac{\pi \sigma^2}{L^2}
- \sum_{{\bf n}\neq {\bf 0}}
\frac{1}{|{\bf n}|}\textrm{erfc}(\frac{L}{\sigma}|{\bf n}|)
\label{MadelungApproximant}
\end{eqnarray}
where ${\bf n}=(i,j,k)$ denotes an integer vector.
Figure \ref{MadelungCalculation} illustrates the rapid
convergence of the Madelung constant calculated from 
Eq. \ref{MadelungApproximant} for a cubic cell.
This expression converges considerably faster than the expression 
frequently found in the literature:
\begin{eqnarray}
\alpha_0 & \approx & \frac{1}{L^2}\sum_{{\bf g}\neq {\bf 0}}
\frac{4\pi}{g^2}e^{-\sigma^2g^2/4} - \frac{2L}{\sqrt{\pi}\sigma}\nonumber \\
&& 
- \sum_{{\bf n}\neq {\bf 0}}
\frac{1}{|{\bf n}|}\textrm{erfc}(\frac{L}{\sigma}|{\bf n}|).
\label{MadelungApproximant2}
\end{eqnarray}
Although a similar procedure can be applied 
without additional difficulty for any dimensionality,
we draw attention to the fact
that in two dimensions, $\alpha_{\sigma/L}$ is not equal to
the Madelung constant $\alpha$ in the limit $\sigma/L \to 0$, due to the 
logarithmic divergence of the potential.
For a more complete discussion of the two-dimensional case,
we refer the reader to the study of Cichocki and Felderhof \cite{CichockiFelderhof1989}. 
As a final remark, we note that the one-dimensional Madelung constant
can be determined analytically from the relation:
\begin{equation}
\sum_{n=1}^{+\infty} \frac{1}{n^2} = \zeta(2) = \frac{\pi^2}{6},
\end{equation}
where $\zeta$ stands for the Riemann zeta function.

\begin{table*}
\scriptsize
\begin{center}
\begin{tabular*}{\textwidth}{@{\extracolsep{\fill}}ll@{\extracolsep{\fill}}ll@{\extracolsep{\fill}}ll@{\extracolsep{\fill}}}
\hline
\multicolumn{2}{l}{3 D} & \multicolumn{2}{l}{2 D} & \multicolumn{2}{l}{1 D} \\
lattice & $\alpha_0$ & lattice & $\alpha$ & lattice & $\alpha_0$ \\ 
\hline
& & & & & \\
cubic & 2.837 297 479 & squared & 2.621 065 852 & linear \, \, \, \, \, \, \, & $-\pi/3$ \\
body-centered  & 3.639 233 449 & hexagonal & 2.786 075 893 &  &  \\
face-centered & 4.584 862 074  &  &  & & \\
 & & & & & \\
\hline
 & & & & & \\
\multicolumn{2}{l}{$v_{\sigma}(r)=\frac{1}{r}\textrm{erf}(\frac{r}{\sigma})$ } & \multicolumn{2}{l}{$v_\sigma(r)=-\ln(\frac{r^2}{\sigma^2}) + \textrm{Ei}( - \frac{r^2}{\sigma^2} )$} & \multicolumn{2}{l}{  $v_\sigma(z)=-2 \pi (z \textrm{erf}(\frac{z}{\sigma}) + \frac{\sigma}{\sqrt{\pi}}e^{ - \frac{z^2}{\sigma^2}})$} \\
\multicolumn{2}{l}{ $v'_{\sigma,L}(r)= \frac{1}{V}\sum_{\bf g \neq \bf 0} \frac{4\pi}{g^2}e^{-\sigma^2 g^2/4+i{\bf g \cdot r}}$ } & \multicolumn{2}{l}{ $v'_{\sigma,L}(r)= \frac{1}{S}\sum_{\bf g \neq \bf 0} \frac{4\pi}{g^2}e^{-\sigma^2 g^2/4+i{\bf g \cdot r}}$ } & \multicolumn{2}{l}{ $v'_{\sigma,L}(z)=\frac{1}{L}\sum_{g \neq 0} \frac{4\pi}{g^2}e^{-\sigma^2 g^2/4+ig \cdot z}$} \\
 & & & & & \\
\hline
 & & & & & \\
\multicolumn{2}{l}{ $v_{\sigma}(0)=\frac{2}{\sqrt{\pi}\sigma}$} & \multicolumn{2}{l}{ $v_\sigma(0)=\gamma$ } & \multicolumn{2}{l}{ $v_\sigma(0)=-2\sqrt{\pi}\sigma$} \\
\multicolumn{2}{l}{ $v'_{\sigma,L}(0)=\frac{2}{\sqrt{\pi}\sigma} - \frac{\alpha_0}{L} + \frac{\pi \sigma^2}{L^3}+...$} & 
\multicolumn{2}{l}{ $v'_{\sigma,L}(0)=\ln(\frac{L^2}{\sigma^2}) - \alpha + \gamma +  \frac{\pi \sigma^2}{L^2}$} &
 \multicolumn{2}{l}{ $v'_{\sigma,L}(0)=-L\alpha_0 - 2 \sqrt{\pi}\sigma+\frac{\pi \sigma^2}{L}+...$} \\
& & & $+...$& & \\
\multicolumn{2}{l}{ $\alpha_{\sigma/L}=\alpha_0 - \frac{\pi \sigma^2}{L^2}+...$} & 
\multicolumn{2}{l}{ $\alpha_{\sigma/L}=-\ln(\frac{L^2}{\sigma^2})  + \alpha - \frac{\pi \sigma^2}{L^2}+...$} & \multicolumn{2}{l}{ $\alpha_{\sigma/L}(0)=\alpha_0 - \frac{\pi \sigma^2}{L}+...$} \\
& & & & & \\
\hline
\end{tabular*}
\end{center}
\caption{Madelung constants in one, two, and three dimensions
computed using the procedure described in Appendix \ref{MadelungAppendix},
along with the quantities used in the calculation. 
Ei denotes the exponential integral and $\gamma=$ 0.577 215 665 is the Euler constant. 
\label{MadelungTable}}
\end{table*}

\section{Performance of the Multipole-expansion Method}

\begin{figure}
\includegraphics[width=8.5cm]{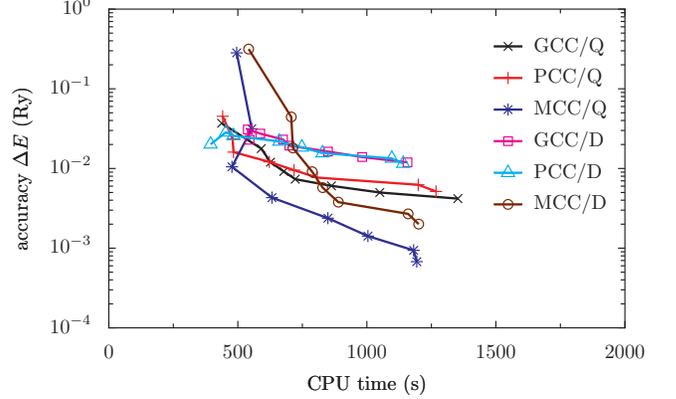}
\caption{Accuracy of the total energy of a pyridazine cation as a function of
computational time using the PCC, GCC, and MCC schemes
for cell sizes in the range
12-19 bohr. The labels
$D$ (dipole) and $Q$ (quadrupole) indicate the order
of the multipole expansion. For each scheme the corrective potential
is updated every five SCF iterations.}
\label{MultipoleAccuracy}
\end{figure}

\label{MultipoleAppendix}

The performance of the multipole-expansion adaptation of
the DCC scheme---the multipole-countercharge (MCC) correction---for
a pyridazine cation is compared to 
that of the PCC and GCC schemes in Figure \ref{MultipoleAccuracy}.
The size of the calculation cell ranges from 12 to 19 bohr.
The parameters used in these calculations are those detailed in 
Sec. \ref{DCCApplicationSection}. Note the good performance of the
MCC approach, which improves the energy accuracy by almost one
order of magnitude in comparison with the PCC and GCC schemes for cell sizes above 17 bohr.

\end{document}